\documentstyle[galley,grl,epsfig,amsmath]{agu2001}
\begin{document}
\title{Finite size scaling in the solar wind magnetic field energy density as seen by WIND}
\author{B. Hnat, \altaffilmark{1} S.C. Chapman \altaffilmark{1}, G. Rowlands
\altaffilmark{1}, N.W. Watkins \altaffilmark{2} and W. M. Farrell \altaffilmark{3}}
\altaffiltext{1}{Space and Astrophysics Group, University of Warwick Coventry, CV4 7AJ, UK}
\altaffiltext{2}{British Antarctic Survey, High Cross, Madingley Road,
Cambridge, UK}
\altaffiltext{3}{NASA Goddard Space Flight Center, Greenbelt MD, USA }

\begin{abstract}
Statistical properties of the interplanetary magnetic field
fluctuations can provide an important insight into the solar wind turbulent cascade. Recently, analysis of the Probability Density Functions (PDF)
of the velocity and magnetic field fluctuations has shown that these exhibit
non-Gaussian properties on small time scales while large scale features appear to be uncorrelated.
Here we apply the finite size scaling technique to explore the scaling of
the magnetic field energy density fluctuations as seen by WIND. We find a
single scaling sufficient to collapse the curves over the entire
investigated range. The rescaled PDF follow a non Gaussian distribution with
asymptotic behavior well described by the Gamma distribution arising from
a finite range L\'{e}vy walk. Such mono scaling suggests that
a Fokker-Planck approach can be applied to study the PDF dynamics.
These results strongly suggest the existence of a common, nonlinear process on
the time scale up to $~26$ hours. 
\end{abstract}

\begin{article}
\section{Introduction}
Statistical properties of the interplanetary magnetic field (IMF)
fluctuations are a topic of considerable interest in space research.
The subject is closely related to energy transport and acceleration
processes in the solar wind \citep{cytu,burlaga,goldstein,zelenyi}.
An approach is to consider MHD turbulence as the process responsible for
observed statistical features of the IMF fluctuations such as intermittency
\citep{cytu,burlaga}.
Single point measurements can not uniquely determine the existence of turbulence
\citep{frisch} and ideally one needs to construct a structure function from a
range of spatial locations in the fluid. However, data taken over long intervals
in the solar wind is routinely single point and can yield  strongly suggestive,
if not definitive results, and it is this type of data that we treat here.

A hallmark of statistical intermittency is the presence of large deviations
from the average value on different scales. An increased probability of finding
such large deviations is manifested in the departure of the PDF from
Gaussian toward a leptokurtic distribution \citep{bohr,sorn} (see also
\citep{vanatta}).
It is well established that such intermittency can be caused by the fluctuations
in the rate of energy transfer of the turbulent cascade \citep{cast1} as
suggested in Kolmogorov's 1962 theory \citep{k62}. Following Kolmogorov's ideas
Castaing proposed an empirical model for the PDF of velocity differences
\citep{cast2}.
Given that: 1) for constant energy transfer rate $\epsilon$, the fluctuating
quantity has a Gaussian distribution and 2) the width of the Gaussian has a
log-normal distribution it was shown that the resultant Castaing distribution
gives a good fit to the velocity difference PDFs in laboratory fluid turbulence
experiments. In the case of the solar wind, one can obtain a good fit to the
PDF of velocity and magnetic field fluctuations by this method 
\citep{valvo1,valvo2} or can consider the fluctuations about the mean value
\citep{padhye},but as we shall see this technique can not be applied over the
entire dynamic range of the magnetic field.
Fitting a curve to a real PDF over its full range is rather difficult
due to the large statistical errors present for the less frequently occurring
large fluctuations.
In practice one has to discard significant amount of data in these tails of 
the PDFs and work only around the center of the distribution curves.
For example, in \citet{valvo1} all events larger then $3$ times the standard
deviation of the original sample were neglected.
The PDF rescaling method used in this paper is free of that limitation
as we do not fit a specific distribution to the data. Instead, we extract the
scaling properties of the fluctuations directly from the data. Importantly the
scaling for the entire dynamic range of the investigated PDFs is obtained from
the statistics of the smallest fluctuations.
The scaling is thus derived using points in the center of the distribution (with
very small statistical errors) and then applied to the PDFs as a whole.

We find that the PDF of fluctuations in magnetic energy density is mono scaling.
The corresponding collapse of the rescaled PDF curves reveals the time scales
over which similar physical processes occur and also confirms any correlations
suggested by the inverse power-law form of the power spectra.
Moreover, it quantifies the asymptotic behavior of the distribution of the
fluctuations, which is essential for constraining turbulence models
\citep{bohr}. Given the scaling exponent we can then use a Fokker-Planck
approach to model the statistical behavior of the time series.

The WIND solar wind magnetic field \citep{lepping} dataset \citep{nick1,freem}
comprises over $1$ million, $46$ second averaged samples from January 1995 to December 1998 inclusive, calculated from the WIND spacecraft key parameter database. The selection criteria for solar wind data was given by the
component of the spacecraft position vector along the Earth-Sun line, $X>0$,
and the vector magnitude, $R >30$ RE. The data set includes intervals of both
slow and fast speed streams.
\begin{figure}
\epsfsize=0.48\textwidth 
\centerline{
\leavevmode\epsffile{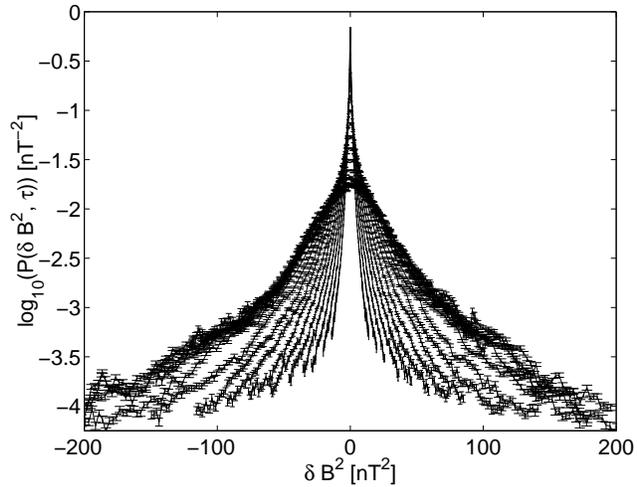}}
\caption{Unscaled PDFs of the IMF energy density fluctuations. Time lag
$\tau=2^k \times 46$s, where $k=0,1,2,..,17$. Standard deviation increases with
$\tau$. Error bars represent $3 \sigma$ intervals assuming Gaussian 
distribution of points within each bin.}
\label{fig1}
\end{figure}
Similarly to other satellite measurements short gaps in WIND data file were
present. To minimize the errors caused by such incomplete measurements we
omitted any intervals where the gap was larger than $15\%$.
The original data were not averaged nor detrended.
The data is not evenly sampled but there are two dominant sampling
frequencies: $1/46$ Hz and $1/92$ Hz. We use sampling frequency $f_s$ of $1/46$
as our base and treat other temporal resolutions as gaps when the accuracy
requires it ($\tau \leq 92$ seconds).
\begin{figure}
\epsfsize=0.48\textwidth 
\centerline{
\leavevmode\epsffile{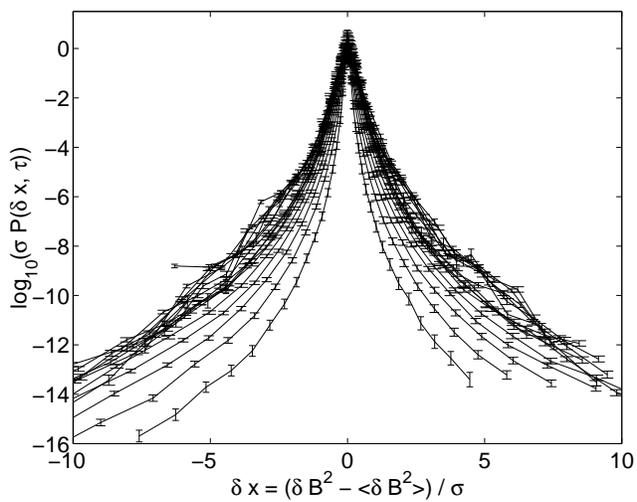}}
\caption{PDFs rescaled with respect to their standard deviation $\sigma(\tau)$.
Shown curves correspond to the scaling region from figure 3 i.e., $\tau$ between
$46$ seconds and $26$ hours. Rescaled PDFs show no satisfactory collapse. Error
bars as in figure 1.}
\label{fig2}
\end{figure}

\section{Scaling of the fluctuation PDF }
Differencing techniques date back to the 1920s when Richardson conveyed the 
image of the turbulent cascade and proposed the velocity difference as an ideal
field for statistical studies \citep{frisch}.
In MHD turbulence the differencing method can also be applied to magnetic field
magnitude measurements. Here, however, we will first investigate the scaling
properties of the magnetic energy density fluctuations given by $B^2(t)$.
Let $B(t)$ represent the time series of the magnetic field magnitude and
$B^2(t)$ its corresponding energy density. A set of time series $\delta
B^2(t,\tau)$ is obtained for each value of time lag $\tau$:
\begin{equation}
\delta B^2(t,\tau)=B^2(t+\tau)-B^2(t)
\label{eq1}
\end{equation}
The PDF $P(\delta B^2,\tau)$ is then calculated for each time series
$\delta B^2(t,\tau)$. Figure \ref{fig1} shows the resultant unscaled PDFs of
the magnetic field energy density fluctuations for different values of the time
lag $\tau=46 \times 2^N$ seconds where $N=0,1,2,...,17$ giving a range of time
scales between $46$ seconds and about $70$ days. We plot the PDF for all
values of $\delta B^2$ for which there are at least $150$ samples per bin.

A generic one parameter rescaling method (the finite size scaling approach,
see e.g. \citet{sorn,diff,montegna}) is applied to these PDFs. 
We first assume that the distributions can be described by a stable, symmetric
law (a Gaussian or L\'{e}vy \citep{bardou}) within a finite scaling range.
We use the peaks of the PDFs to obtain the scaling exponent $s$. This is
important as the peaks are the most accurate parts of the distributions.
Figure \ref{fig2} shows these PDFs rescaled with respect to their 
$\sigma(\tau)$ (see e.g. \citep{cast2,vanatta}) and we see that these also do
not collapse onto a single curve.

\begin{figure}
\epsfsize=0.48\textwidth 
\centerline{
\leavevmode\epsffile{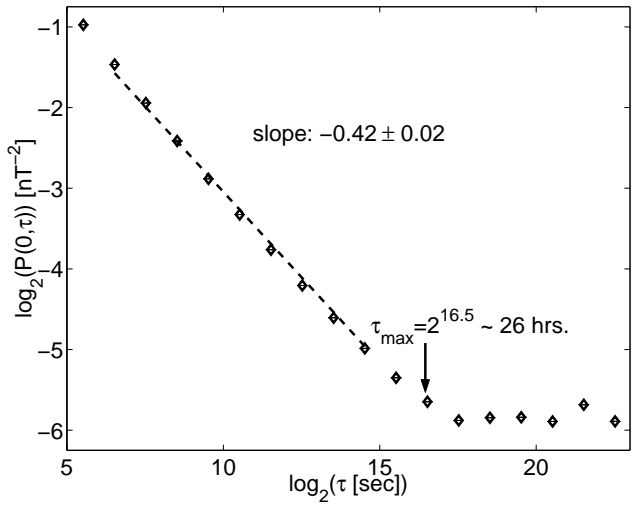}}
\caption{Scaling of peaks $P(0,\tau)$ of the IMF energy density fluctuations
PDFs. Slope of the line is fitted with 95\% confidence interval. Scaling
extends from $\tau=46$ seconds to about $\tau=26$ hours. Error bars as in figure 1.}
\label{fig3}
\end{figure}

\begin{figure}
\epsfsize=0.48\textwidth 
\centerline{
\leavevmode\epsffile{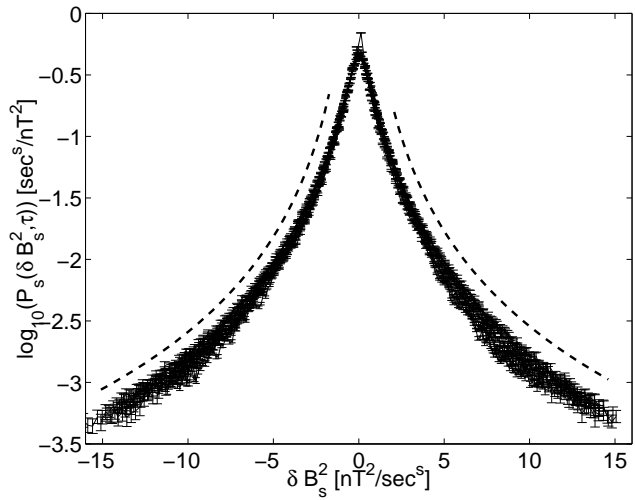}}
\caption{One parameter rescaling of the magnetic field energy density PDFs.
The curves shown correspond to $\tau$ between $46$ seconds and $26$ hours.
Dashed lines represent the Gamma distribution of the large fluctuations
for the PDF of $\tau=2^4$. Vertical offset was added for clarity. Error bars as
in figure 1.}
\label{fig4}
\end{figure}

Figure \ref{fig3} shows the peaks $P(0,\tau)$ of the unscaled PDFs (shown in
Fig. \ref{fig1}) plotted versus $\tau$ on log-log axes. We see that there is
a range of $\tau$ up to $~26$ hours for which $P(0,\tau)$ is well described by
a power law $\tau^{-s}$ with index $s=0.42 \pm 0.02$. This break in the scaling
at $\tau \sim 26$ hours is consistent with that found previously (see e.g
\citet{burlaga,valvo1,valvo2}). Within this range we now attempt to
collapse the entire unscaled PDF shown in Fig. \ref{fig1} onto a single master
curve using the following change of variables:
\begin{equation}
P(\delta B^2,\tau)=\tau^{-s} P_s(\delta B^2 \tau^{-s},\tau)
\label{eq2}
\end{equation}
If the initial assumption of the stable and symmetric distributions was correct
a single parameter rescaling, given by equation (\ref{eq2}), for a mono-fractal
process, would give a perfect collapse of PDFs on all scales. A self-similar
Brownian walk is a good example of the process where such collapse can be
observed (see e.g. \citet{sorn}).
Figure \ref{fig4} shows the result of the above one parameter rescaling applied to the unscaled PDF of IMF energy density fluctuations shown in Fig \ref{fig1}.
We see that the rescaling procedure (\ref{eq2}) using the value of the exponent $s$ of the peaks $P(0,\tau)$, as shown in Fig \ref{fig3}, gives good collapse
of the curves onto a single common functional form for the entire range of the
data. 

The successful rescaling of the PDFs now reveals some key features of the
processes governing the magnetic field energy density fluctuations.
In particular, as we will show below, one can quantify an asymptotic behavior
of the rescaled PDFs. A single parameter rescaling strongly suggests that the
resultant distribution is stable and as such should converge to Gaussian or
L\'{e}vy distribution for large fluctuations \citep{sorn}.
The leptokurtic character of the PDFs is still present after the rescaling
indicating that the L\'{e}vy distribution is more appropriate to model the
functional form of the curves.
For a system where $B^2$ has an upper, physical or instrumental, limit such a
distribution is truncated.
Importantly, truncated L\'{e}vy distributions are still 1) stable and 2) go
over to Gaussian for sufficiently large time lag $\tau$ \citep{sorn}.
Calculation of the Hurst exponent for the $B^2$ time series, by direct examination of the $\sigma^2\propto\tau^{2H}$ relation as well as the 
``growth of range" algorithm, confirms such a limit does exist - we find a mean
exponent $H=0.28 \pm 0.02$ up to $~26$ hours beyond which the variance of the
signal saturates.
We then fit a truncated L\'{e}vy distribution to the large $\delta B^2$
asymptotes of the rescaled PDFs. This Gamma distribution is of the form
\citep[see][]{sorn}:
\begin{equation}
P(\delta B^2)=C_{\pm} exp(-\frac{|\delta B^2|}{\Delta})|\delta B^2|^{-(1+\frac{1}{\alpha})}
\label{eq4}
\end{equation}
Two dashed lines on Fig. \ref{fig4} show (displaced for clarity) the Gamma
distribution of the large energy density fluctuations. This best fit of
(\ref{eq4}) to each of the tails of the rescaled PDFs is obtained for the power
law index $\alpha \sim 0.66$ and parameters $C_-=0.9$, $C_+=1$ and $\Delta=100$.
The maximum time scale of about $26$ hours beyond which the scaling
no longer holds is in good agreement with previous results
\citep{valvo1,valvo2}, (see also \citet{burlaga} and references therein).

\begin{figure}
\epsfsize=0.48\textwidth 
\centerline{
\leavevmode\epsffile{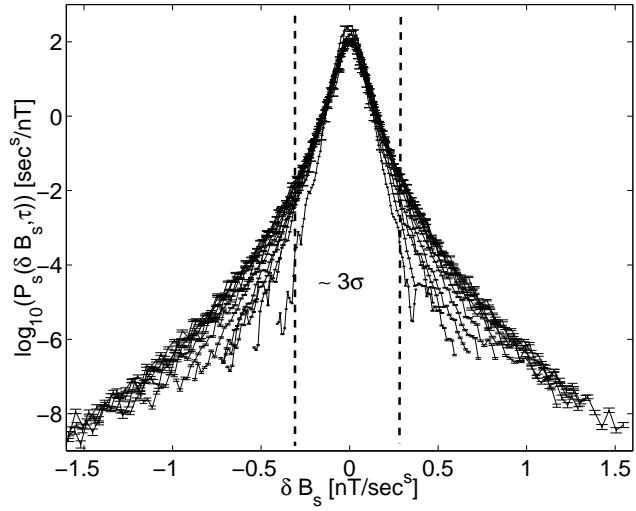}}
\caption{One parameter rescaling of PDFs of IMF fluctuations. Shown curves
correspond to $\tau$ between $46$ seconds and $26$ hours. Note the satisfactory collapse to about $3$ standard deviations of the original sample
and poor collapse of the tails. Error bars as in figure 1.}
\label{fig5}
\end{figure}

It has been found previously \citep{burlaga} that the magnetic field magnitude
fluctuations are not self-similar but rather multi-fractal.
For such processes the scaling derived from $P(0,\tau)$ would not be expected
to rescale the entire PDF as above. To verify this we repeated the
rescaling procedure for magnetic field magnitude differences $\delta B(t,\tau)=B(t+\tau)-B(t)$.
Figure \ref{fig5} shows the result of one parameter rescaling applied to the
PDFs of the magnetic field magnitude fluctuations. The scaling procedure is satisfactory only up to $3$ standard deviations of the original sample. This confirms the results of \citet{valvo1,valvo2} where a two parameter Castaing fit to values within $3$ standard deviations of the original sample yields scaling
in one parameter and weak variation in the other. Attempts to improve the
collapse by using information in the tails (values $|\delta B| > 3 \sigma$)
would introduce a significant error in the estimation of the scaling exponent
$s$.
However, the difference in scaling between the PDFs of $\delta B^2$
and $\delta B$ seen here may simply arise from the long range correlations
present in the magnetic field magnitude measurements. This follows since:
\begin{equation}
(\delta B)^2=B^2(t+\tau)-B^2(t)-2(B(t)B(t+\tau)-B^2(t))
\label{eq6}
\end{equation}
The first two terms, $B^2(t+\tau)-B^2(t)$ are the fluctuation of the
magnetic energy density and as we have seen this quantity is mono-scaling.
The last term, which includes $B(t)B(t+\tau)$, will however be significant
when the  autocorrelation function of the IMF magnitude time series is non
negligible.

The Fokker-Planck equation provides an important link between statistical
studies and the dynamical approach expressed by the Langevin equation
\citep{sorn}. The mono-scaling of the magnetic field energy density PDF,
together with the finite value of the samples' variance, indicates that a
Fokker-Planck approach can be used to study the dynamics of the unscaled PDF
$P(\delta B^2,t)$ in time and with respect to the coordinate $\delta B^2$
\citep{kampen}.
Recently a fractional Fokker-Planck equation has been obtained that can be used
to treat L\'{e}vy walk PDF dynamics \citep{lovejoy}. Alternatively, the
anomalous diffusion of the PDF, consistent with truncated
L\'{e}vy walk, can be obtained by introduction of a functional dependence of the
diffusion coefficient on the ``spatial" coordinate (in our case $\delta B^2$).
One can then write a Fokker-Planck equation for the evolution of the PDF as:
\begin{equation}
\frac{\partial{P}}{\partial{t}}=
\nabla_{\beta} (P \gamma (\delta B^2))+\nabla_{\beta}^2 D P 
\label{eq7}
\end{equation}
where $P \equiv P(\delta B^2,t)$ is a distribution function, $\gamma$ is
called the friction coefficient and $D \equiv D(\delta B^2)$ is a diffusion
coefficient which in this case depends on the magnetic energy density
fluctuations. The differential operator $\nabla_{\beta}=$ should be understood
as:
\begin{equation}
\nabla_{\beta}=\frac{\partial{}}{\partial{(\delta B^2)}}
\label{eq8}
\end{equation}
Such a functional dependence of the diffusion coefficient on $\delta B^2$ leads
to the anomalous diffusion resulting in the departure from the well known
Einstein law relating variance of the signal and the time increments
\cite{sorn}.

\section{Summary}
In this paper we have applied the generic finite size scaling method to
study the scaling for the magnetic field energy density and magnitude
fluctuations of the IMF using WIND MFI data.
This yields the scaling properties of the solar wind magnetic field energy
density fluctuations that do not assume a particular turbulent model.
Instead, we use the smallest (and best represented statistically)
fluctuations to obtain a single parameter that gives the scaling exponent
sufficient to rescale the $46$ seconds averaged data over its
entire dynamic range in $\delta B^2$.
The simplest explanation suggested by these results is that the IMF energy
density fluctuations are governed by a common physical process that is 
self-similar on the temporal scale up to $\tau \sim 26$ hours.
The rescaled PDF follows a non-Gaussian distribution with an asymptotic behavior
well described by the Gamma distribution arising from a finite range L\'{e}vy walk. Such rescaling also indicates that a Fokker-Planck approach can be used to
study the evolution (in $\delta B^2$ space) of the PDF.
A similar rescaling procedure applied to magnetic field magnitude fluctuations
confirms more complex, perhaps multi-fractal, character of the underlying dynamics in agreement with previous results.

\section{Acknowledgment} 
S. C. Chapman and B. Hnat acknowledge support from the PPARC and G. Rowlands
from the Leverhulme Trust. We thank M. P. Freeman and D. J. Riley
for post processing the WIND data. We thank J. Greenhough for useful
discussions and R.P Lepping for provision of data from the NASA WIND spacecraft.

\end{article}
\end{document}